\documentstyle[aps,prl,multicol,epsfig]{revtex}

\newcommand{\lapprox}{\stackrel{<}{\sim}}
\newcommand{\gapprox}{\stackrel{>}{\sim}}

\begin{document}


\title{{\bf Density-induced BCS to Bose-Einstein crossover}}
\vspace{1cm}
\par
\author{N. Andrenacci$^{(1)}$, A. Perali$^{(1,2)}$, P. Pieri$^{(1)}$,
        and G.C. Strinati$^{(1)}$}
\address{(1) Dipartimento di Matematica e Fisica, Sezione INFM, 
Universit\`{a} di Camerino, I-62032 Camerino, Italy}
\address{(2) Dipartimento di Fisica, Sezione INFM, 
Universit\`{a} di Roma ``La Sapienza'', I-00185 Roma, Italy}

\date{\today}
\maketitle

\begin{abstract}
We investigate the zero-temperature BCS to Bose-Einstein crossover at the
mean-field level, by driving it with the attractive potential \emph{and\/}
the particle density.
We emphasize specifically the role played by the particle density in this 
crossover.
Three different interparticle potentials are considered for the continuum model 
in three spatial dimensions, while both $s$- and $d$-wave solutions are analyzed 
for the attractive (extended) Hubbard model on a two-dimensional square lattice.
For this model the peculiar behavior of the crossover for the $d$-wave solution 
is discussed.
In particular, in the strong-coupling limit when approaching half filling we
evidence the occurrence of strong correlations among antiparallel-spin fermions
belonging to different composite bosons, which give rise to a quasi-long-range
antiferromagnetic order in this limit.
\end{abstract}
\begin{multicols}{2}
\vspace{1cm}

\section{Introduction}

The BCS to Bose-Einstein (BE) crossover has been widely studied in the last several
years,\cite{NSR,Randeria-93,Haussmann,PS-96,Zwerger,Levin} being originally motivated 
by the occurrence of a short coherence length in high-temperature superconductors.
The evolution from large overlapping Cooper pairs (BCS limit) to small nonoverlapping 
bosons (BE limit) has essentially been envisaged by relying on the associated two-body
problem in the three-dimensional case, wherein bound fermion pairs (composite bosons) 
form as soon as the strength of the attractive interparticle potential exceeds a 
threshold.
The emphasis on the role of the interparticle potential has, however, somewhat 
overshadowed the effects of the particle density on the crossover itself, even though 
on physical grounds one would expect both the interparticle potential \emph{and\/} the 
density to play an essential role.
The role played by the density is suggested especially when one analyzes the 
experimental phase diagram of the high-temperature cuprate superconductors in terms of
the BCS-BE crossover, since in this case it would be the (effective) carrier density 
(that is related to the doping level) to drive the system from the vicinity of the BE 
(underdoped) to the BCS (overdoped) limit.\cite{Uemura,footnote-1}

Purpose of this paper is to study the \emph{combined\/} effects of the particle
density and the interparticle potential on the (zero-temperature) BCS-BE crossover,
in order to characterize how physical quantities evolve by varying, in particular, 
the particle density.
To this end, we will set up a ``phase diagram'' in the space of the potential 
strength and of the density, where the locations of the alternative BCS-like, 
crossover-like, and BE-like regions will be identified for several types
of potentials.

Previous work on the BCS-BE crossover has utilized: \emph{(i)\/} a three-dimensional
contact potential for the continuum model \cite{Randeria-93,Haussmann}; \emph{(ii)\/} 
the separable potential introduced by Nozi\`{e}res and Schmitt-Rink (NSR) \cite{NSR} 
for the three-dimensional continuum model, with a characteristic momentum cutoff
$k_{o}$ \cite{PS-96,Levin}; and \emph{(iii)\/} a negative-$U$ Hubbard model on a 
two-dimensional square lattice, with either an on-site \cite{Micnas} or a 
nearest-neighbor \cite{Micnas,Randeria-98} attraction.
In this paper, we shall consider both the continuum model in three spatial dimensions
using three different types of interparticle potentials, and the $s$- and $d$-wave
solutions for the attractive (extended) Hubbard model on a two-dimensional square
lattice.
This will enable us to study the effects of the particle density on the BCS-BE 
crossover in a rather systematic way.

For the continuum case, it turns out that the \emph{finite range\/} of the potential
allows for the occurrence of the density-induced BCS-BE crossover, which is instead 
not possible in the case of a contact (zero-range) potential.
As a consequence, the size of the BCS-like region in the ``phase diagram'' gets
progressively enlarged by increasing the range of the interparticle potential.

For the lattice case, the shape of the ``phase diagram'' and the physical 
interpretation of the alternative regions therein depend markedly on the symmetry
($s$ or $d$) of the pairing.
In particular, for the $d$-wave pairing an increasingly larger range of correlations 
among the composite bosons sets up when approaching half filling, thus establishing 
a tendency toward the formation of a quasi-long-range-ordered antiferromagnetic state.
In addition, the $d$-wave pairing, being associated with an interaction of a finite
range on the lattice, enables the density-induced BCS-BE crossover to occur over a
wider range of the parameters with respect to the $s$-wave pairing, in analogy to 
what found for the continuum case.

All results presented in this paper have been obtained within a zero-temperature
broken-symmetry mean-field approach, which thus appears capable of producing results
that are sensible on physical grounds under widely different physical conditions.

The plan of the paper is as follows. We discuss the three-dimensional continuum case
in Section II and the two-dimensional lattice case in Section III. Section IV gives
our conclusions.

\section{Three-dimensional continuum case}

In this Section, we examine the three-dimensional BCS-BE crossover in the continuum
case for three types of interaction potentials, and determine how the range of the
potential influences the coupling vs density ``phase diagram''.
Specifically, we consider the contact and the separable NSR potentials mentioned in 
the Introduction, plus a non-separable Gaussian potential.

For the three-dimensional continuum case, a \emph{contact potential\/} has often 
been adopted as \emph{the reference model\/} that captures the essence of the 
expected physics of the BCS-BE crossover, as a function of the coupling strength 
for given particle density.
For a contact potential, the analytic solution at the (zero temperature)
mean-field level and with the inclusion of Gaussian fluctuations has been 
determined,\cite{MPS} for \emph{all\/} values of the coupling strength (regularized 
in terms of the scattering length $a$ of the associated two-body problem) and of the 
density (represented in terms of the Fermi wave vector $k_{F}$).
All relevant physical quantities can thus be expressed in terms of the dimensionless
parameter $k_{F} a$, with $k_{F}$ being positive by definition and $a$ changing its
sign as soon as a bound state develops.
For this reason, by keeping $k_{F}$ fixed and varying $a$ from $-\infty$ to $+\infty$
one can pass with continuity from the BCS to the BE regime across the crossover region;
on the contrary, by keeping $a$ fixed and varying $k_{F}$ one is \emph{not\/} able to 
pass from the BCS to the BE regime, since the parameter $k_{F} a$ cannot change its 
sign in this way.

\begin{figure}
\narrowtext
\epsfxsize=8truecm
\epsfbox{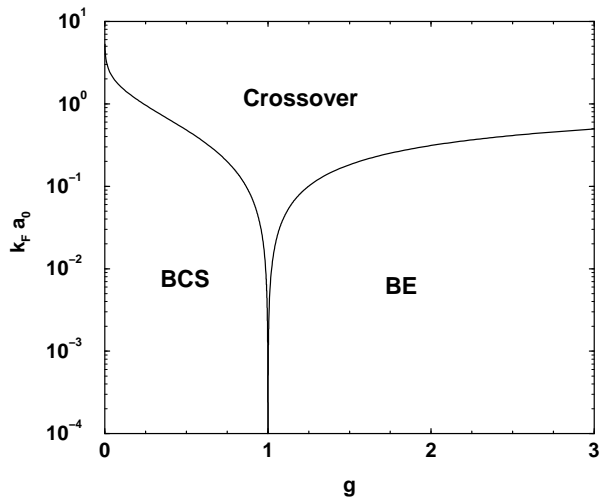}
\caption{Phase diagram ($g,k_Fa_o$) for a contact potential in 
three dimensions, as obtained from the available analytic 
solution (see text).}
\end{figure}
Figure~1 shows the ($g$,$k_{F}a_{o}$) ``phase diagram'' for a contact potential in 
three dimensions, as obtained from the available analytic solution,\cite{MPS} where 
BCS-like, crossover-like, and BE-like regions are identified by drawing the two 
curves corresponding to $k_{F} \xi_{pair} = 2 \pi$ and $k_{F} \xi_{pair} = 1/ \pi$. 
Here, $\xi_{pair}$ represents the correlation length for pairs of opposite-spin
fermions, while the two values ($2 \pi,1/ \pi$) of the parameter $k_{F} \xi_{pair}$ 
characterize, in the order, the lower limit of the BCS-like region (with large 
overlapping Cooper pairs) and the upper limit of the BE-like region (with small 
nonoverlapping bosons), the crossover region being constrained in between.\cite{PS-94}
The (effective) coupling constant in this case is represented by $g=\exp{(a_{o}/a)}$, 
where $a_{o}$ is an arbitrary unit of length.
As anticipated above, it is evident from this ``phase diagram'' that for a contact
potential in three dimensions it is \emph{not\/} possible to cross over from the BCS 
to the BE region by varying the density alone at fixed coupling strength.

In the three-dimensional continuum case, therefore, to examine the density-induced 
BCS-BE crossover a potential with \emph{finite range\/} in real space is required, or 
equivalently, it is necessary to introduce a momentum cutoff $k_{o}$ in momentum space.
In this context, one may utilize the separable potential 
$V({\mathbf k},{\mathbf k}')=V w({\mathbf k}) w({\mathbf k}')$ (between fermions with
opposite spins) adopted in Ref.~\cite{NSR}, with 
$V<0$ and $w({\mathbf k})=(1+(|{\mathbf k}|/k_{o})^{2})^{-1/2}$, and later used by 
some authors.\cite{PS-96,Levin}
Since this potential may yield unphysical results \cite{PS-96} (also because the 
factorization $w({\mathbf k}) w({\mathbf k}')$ is somewhat arbitrary), we have 
considered in addition a non-separable potential 
$V({\mathbf k},{\mathbf k}')=V({\mathbf k}-{\mathbf k}')$, which we have taken for
definiteness of the Gaussian form:

\begin{equation}
V({\mathbf k}-{\mathbf k}') \, = \, V \, 
\exp \left\{ -|{\mathbf k}-{\mathbf k}'|^{2}/k_{o}^{2} \right\}
\label{V}
\end{equation}

\noindent
($V<0$). By doing so, it will also be possible to determine how alternative ways
of introducing an effective range in the interaction  potential affect the boundaries 
of the crossover region in the ($|V|/V_{c},k_{F}/k_{o}$) ``phase diagram'' 
($V_{c}$ being the critical value of $|V|$ for which a bound state appears in the 
two-body problem).
In addition, and contrary to the separable NSR potential, the Gaussian potential 
(\ref{V}) leads to momentum-decoupling effects for small values of $k_{o}$, 
which have recently been proposed as characteristic features of high-temperature
superconductors.\cite{Perali}

For the NSR and Gaussian potentials, an analytic solution for the (zero-temperature) 
BCS-BE crossover at the mean-field level is lacking.
For these potentials, we have thus solved numerically the coupled equations for the 
gap function $\Delta({\mathbf k})$ and the chemical potential $\mu$:

\begin{equation}
\Delta({\mathbf k}) \, = \, - \, \int \! \frac{d{\mathbf k}'}{(2\pi)^{3}} \,
V({\mathbf k},{\mathbf k}') \, \frac{\Delta({\mathbf k}')}{2 E({\mathbf k}')}  \label{Delta}
\end{equation}

\noindent
and

\begin{equation}
n \, = \, \int \! \frac{d{\mathbf k}'}{(2\pi)^{3}} \, 
\left( 1 - \frac{\xi({\mathbf k}')}{E({\mathbf k}')} \right) \, \, ,     \label{mu}
\end{equation}

\noindent
where $\xi({\mathbf k})={\mathbf k}^{2}/(2m)-\mu $ ($m$ being the fermionic mass),
$E({\mathbf k})=\sqrt{\xi({\mathbf k})^{2}+\Delta({\mathbf k})^{2}}$, and $n$ is the
particle density.
We recall that, while for a separable potential the gap function acquires the form
$\Delta({\mathbf k})=\Delta_{o} w({\mathbf k})$, for a non-separable potential
$\Delta({\mathbf k})$ does not follow in general the wave vector dependence of the
potential.
In addition, for the NSR separable potential analytic expressions for $\Delta_{o}$ 
and $\mu$ can be obtained in the two (BCS and BE) limits.

\begin{figure}
\narrowtext
\epsfxsize=8truecm
\epsfbox{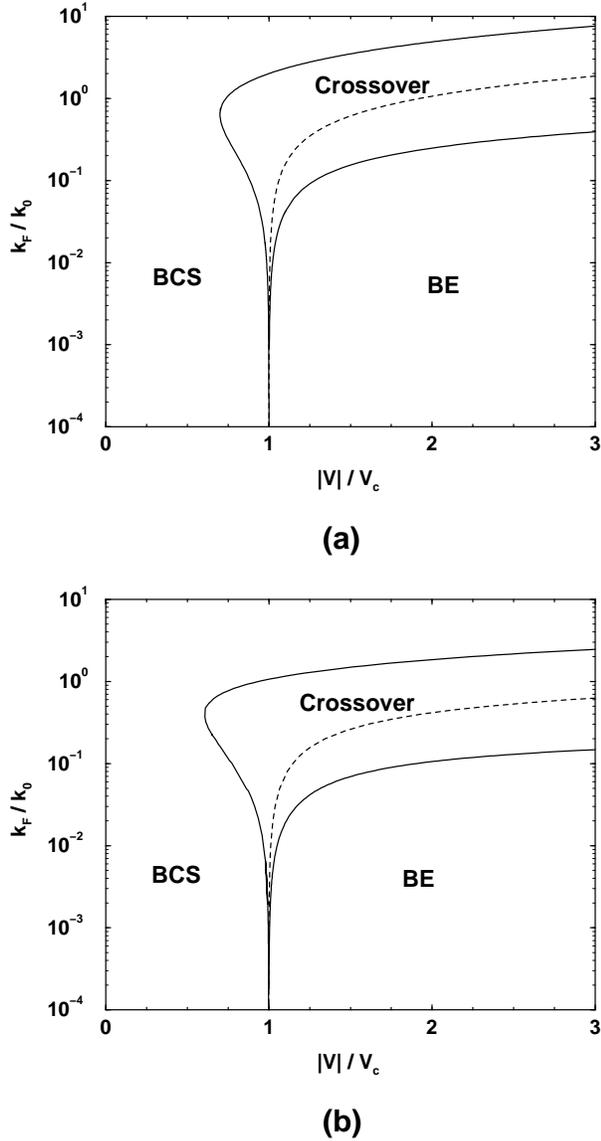}
\vspace{0.4truecm}
\caption{Phase diagram ($|V|/V_{c},k_{F}/k_{o}$) for the (a) NSR and 
(b) Gaussian potentials in three dimensions (see the text for the meaning of 
the different curves).}
\end{figure}

Figure~2 shows the ($|V|/V_{c}$,$k_{F}/k_{o}$) ``phase diagram'' for the (a) NSR 
and (b) Gaussian potentials, where the two characteristic curves
$k_{F} \xi_{pair} = (2 \pi,1/ \pi)$ for each potential have been identified as 
for Fig.~1.
In addition, we have reported in Fig.~2 the curve corresponding to $\mu=0$ (broken 
line) for both potentials.
By comparing Fig.~2 with Fig.~1, we note that the boundary between the BE-like and
the crossover-like regions is not much altered by the introduction of a finite
cutoff $k_{o}$; on the contrary, the boundary between the BCS-like and the
crossover-like regions is drastically modified, with the BCS-like region extending
even to values $|V|>>V_{c}$ for sufficiently high densities.
Note also the reentrant shape of the curve $k_{F} \xi_{pair} = 2 \pi$ at 
$k_{F} \simeq k_{o}$ for both potentials, which makes it possible to cross over from
the BCS to the BE region \emph{by varying the density\/} for fixed $|V|>V_{c}$.

It is further interesting to note that the three curves of Fig.~2a and of Fig.~2b 
depart from the common origin ($|V|/V_{c}=1$,$k_{F}=0$). 
Near this origin, in fact, $k_{F}<<k_{o}$ and the gap 
$\Delta_{o} = \Delta({\mathbf k}=0)$ on the BCS side is proportional to 
$(k_{F}^{2}/2m) \exp\{\pi/(2k_{F}a)\}$, as given by the solution for a contact 
potential.\cite{Haussmann,MPS}
Since $\xi_{pair} \propto k_{F}/\Delta_{o}$ in the BCS limit,\cite{PS-94} keeping the 
product $k_{F} \xi_{pair} \propto \exp\{-\pi/(2k_{F}a)\}=\mathrm{constant}$, requires
$a \rightarrow - \infty$ when $k_{F} \rightarrow 0$.
This, in turn, implies that the curves corresponding to 
$k_{F} \xi_{pair} \gapprox 2\pi$ depart from the point $|V|=V_{c}$ on the $k_{F}=0$ 
axis.
On the BE side, on the other hand, when $k_{F}<<k_{o}$ (and $|V|>V_{c}$), 
$\xi_{pair}$ coincides with the bound-state radius $r_{o}$;\cite{PS-94} keeping thus 
constant the product $k_{F} \xi_{pair}=k_{F} r_{o}$ when $k_{F} \rightarrow 0$ implies
$r_{o} \rightarrow \infty$, i.e., $|V| \rightarrow V_{c}$.\cite{footnote-2}

The reentrant shape of the curve $k_{F} \xi_{pair}=2 \pi$ in Fig.~2 at 
$k_{F} \simeq k_{o}$, too, can be understood by simple analytic arguments as follows.
When $k_{F}<<k_{o}$, the expression $\Delta_{o}/\mu \propto \exp\{\pi/(2k_{F}a)\}$,
which is valid on the BCS side for a contact potential, can again be used.
At fixed value of $|V|<V_{c}$ (such that the scattering length $a$ is negative),
$\Delta_{o}/\mu$ vanishes when $k_{F} \rightarrow 0$.
In this way, one approaches the (weak-coupling) BCS limit for decreasing $k_{F}$
at fixed $a<0$.
When $k_{F}>>k_{o}$, on the other hand, it is necessary to distinguish the NSR
from the Gaussian potential.
For the NSR potential, the value of $\Delta(k_{F})=\Delta_{o} w(k_{F})$ can be 
obtained analytically in the BCS limit (and $k_{F}>>k_{o}$), in the form

\begin{equation}
\frac{\Delta(k_{F})}{\mu} \, \approx 8 \, \exp \left\{ 
                \frac{1}{N_{o}V(k_{F},k_{F})}  \right\}       \label{Delta/mu}
\end{equation}

\noindent
where $N_{o}=m k_{F}/(2 \pi^{2})$ is the density of states at the Fermi level (per
spin component) and $V(k_{F},k_{F}) \simeq (k_{o}/k_{F})^{2} V$.
In this case, the decrease of $|V(k_{F},k_{F})|$ for increasing $k_{F}/k_{o}$ 
overcomes in Eq.~(\ref{Delta/mu}) the increase of $N_{o}$, and drives the system 
toward the BCS (weak-coupling) limit.
For the non-separable Gaussian potential, on the other hand, it is the effective 
reduction of the density of states and not the decrease of the potential strength 
to drive the system toward the BCS limit for increasing $k_{F}/k_{o}$ at given $V$.
To verify this statement, we recall that in the BCS limit $E({\mathbf k}')$ in
Eq.~(\ref{Delta}) is strongly peaked about $k_{F}$.
When $k_{F}>>k_{o}$, $\Delta({\mathbf k})$ is thus also strongly peaked about $k_{F}$
owing to the form (\ref{V}) of the potential.
In this way, for given value of ${\mathbf k}$, the integral over ${\mathbf k}'$
in Eq.~(\ref{Delta}) extends effectively over a sphere centered about ${\mathbf k}$
with radius of the order $k_{o}$.
The relevant density of states gets thus reduced from the value $N_{o}$ by a
geometrical factor $R$ of the order $(4\pi k_{o}^{3}/3)/(8\pi k_{F}^{2} k_{o})$,
which represents the ratio of the effective volume of integration and the BCS spherical
shell of width $2 k_{o}$.
We then obtain in the BCS limit for the Gaussian potential when $k_{F}>>k_{o}$:

\begin{equation}
\Delta(k_{F}) \, \propto \, \frac{k_{o} k_{F}}{m} \, 
           \exp \left\{ \frac{1}{R N_{o} V} \right\} \, .   
\label{Delta-prop}
\end{equation}

\noindent
In this case, it is thus the decrease of $R$ for increasing $k_{F}$ to drive the
system toward the BCS (weak-coupling) limit for a given value of $V$.

In summary, we have shown that, although the qualitative behavior of the curves 
corresponding to $k_{F} \xi_{pair} = 2\pi$ in Figs.~2a and 2b looks similar, the 
physical mechanism behind them appears to be quite different.
For the separable NSR potential the increase of $k_{F}/k_{o}$ results into a reduction 
of the interaction strength, while for the non-separable Gaussian potential it results
into a reduction of the relevant density of states.
These two alternative effects allow the density-induced BCS-BE crossover
to occur in the two cases, respectively.

The curves of Fig.~2 have been drawn cautiously, up to values of $|V|/V_{c}$ for
which the condition $|\mu|<<k_{o}^{2}/(2m)$ is satisfied.
This condition avoids, in fact, instabilities of the system in the bosonic limit,
which unavoidably occur when a fermionic potential with \emph{finite\/} momentum 
range $k_{o}$ is considered, and are due to the boson-boson effective interaction 
potential having a dominant attractive part in this case.\cite{PS-96,Pi-S-98}
We have verified, in particular, for the Gaussian potential that the bosonic
chemical potential $\mu_{B} = 2 \mu + \epsilon_{o}$ (where $\epsilon_{o}$ is the
bound-state energy of the associated two-body problem) \cite{Haussmann,PS-96} becomes
negative when $|\mu| \gapprox k_{o}^{2}/m$, a behavior which can be attributed to
an overall attraction among the composite bosons (with the compressibility being,
however, still positive).  
In this context, we mention also that the existence of a competition between pair 
and quartet condensation in a Fermi liquid with a finite-range attraction has 
recently been investigated.\cite{Ropke}

A related instability toward phase separation (with the compressibility becoming
instead negative) has been generically pointed out for the attractive (extended) 
Hubbard model on a two-dimensional square lattice,\cite{Micnas} which we pass now 
to examine in detail in the context of the BCS-BE crossover.

\section{Two-dimensional lattice case}

In this Section, we examine the BCS-BE crossover for a two-dimensional attractive 
Hubbard model, again addressing specifically the role played by the particle density 
in driving this crossover.
To consider the $d$-wave besides the $s$-wave solution, we take the fermionic
potential to contain an inter-site besides an on-site term.
In addition, for the $d$-wave solution we adopt two different single-particle 
dispersions, in order to mimic the low-energy electronic band structure observed 
for the cuprates in different doping ranges,\cite{Ino,Saini} as discussed in more 
detail below.
We recall that the issue of the $d$-wave symmetry in the context of the BCS-BE
crossover has been discussed briefly in Ref.~\cite{Levin-98} and more extensively in
Refs.~\cite{Micnas} and \cite{Hertog}.

For this model we thus take:

\begin{equation}
V({\mathbf k},{\mathbf k}') \, = \, U \, + \, 2 \, V \, 
           (\cos(k_{x}-k'_{x}) \, + \,\cos(k_{y}-k'_{y}))   
\label{V-Hubbard}
\end{equation}

\noindent
where $V \le 0$.
In particular, when $V=0$ we consider an on-site attraction $U<0$, with the hopping 
in the kinetic term of the fermionic Hamiltonian limited to nearest-neighbor sites 
(which corresponds to the ordinary negative-$U$ Hubbard model \cite{Micnas});
when $V<0$ we consider instead an on-site repulsion $U>0$, with the hopping in the 
kinetic term of the fermionic Hamiltonian either limited to nearest-neighbor sites 
or ranging over second- and third-neighbor sites (which corresponds to an extended 
attractive Hubbard model, with an on-site repulsion \cite{Dagotto}).
Recall that the term proportional to $V$ in Eq.~(\ref{V-Hubbard}) is associated with
an attraction between opposite-spin fermions on neighboring sites in the square 
lattice.

\subsection{$s$-wave solution}

The negative-$U$ Hubbard model (obtained by taking $V=0$ and $U<0$ in 
Eq.~(\ref{V-Hubbard})) plays on the lattice an analogous role to the contact 
potential in the continuum case.
For this model, the only nontrivial solution to the gap equation (\ref{Delta})
(with the integration over the wave vector being now limited to the two-dimensional
Brillouin zone (BZ)) has $s$-wave symmetry ($\Delta({\mathbf k})=\Delta_{o}$), and the
dispersion relation $\xi({\mathbf k}) = -2t(\cos k_{x}+\cos k_{y}) + nU/2 - \mu$
($t>0$) contains the Hartree shift $nU/2$.\cite{footnote-3}
The corresponding ($|U|/t,n$) ``phase diagram'' is shown in Fig.~3a, with the
boundaries between the alternative (BCS, crossover, and BE) regions identified 
like for the continuum case.
Note that the critical value $U_{c}$ for which a bound state appears in the
two-body ($n=0$) problem consistently vanishes for the $s$-wave solution in
two dimensions.
Note also that the reentrant shape of the curve $k_{F} \xi_{pair} = 2 \pi$
(as well as of the curve $k_{F} \xi_{pair} = 1/ \pi$) stems now from the fact
that the van Hove singularity of the density of states is approached when
$n$ tends to 1 (half filling).
(Owing to the symmetry of the density of states about half filling, the diagram
of Fig.~3a is also symmetric about half filling.)
Note finally that the crossover from the BCS to the BE region by varying the
density for fixed $|U|$ is possible only for $|U| \lapprox 2.4 t$, the BE and 
crossover regions being depressed in this case to extremely small values of $n$.

\begin{figure}
\narrowtext
\epsfxsize=8truecm
\epsfbox{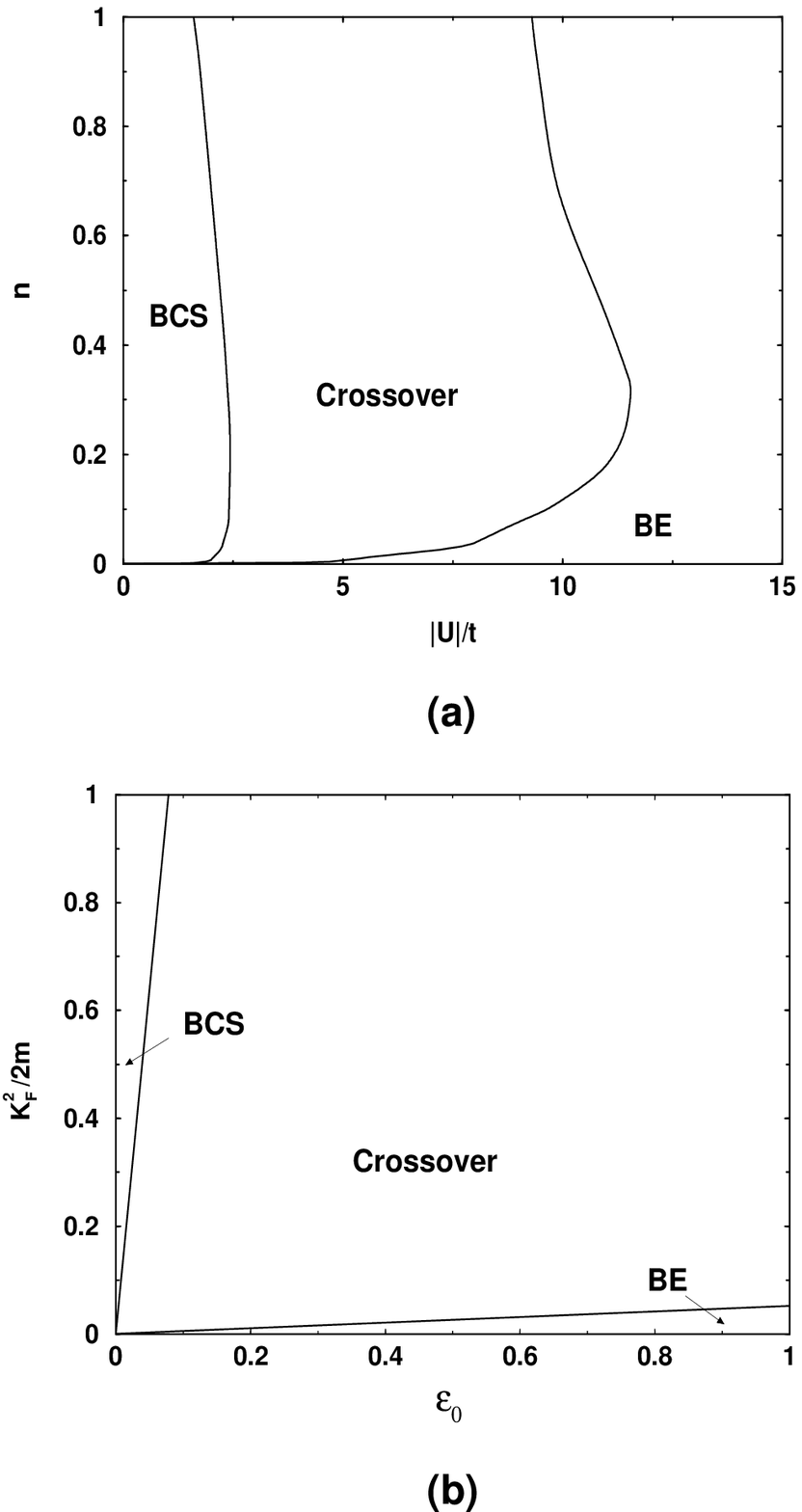}
\vspace{.4truecm}
\caption{(a) Phase diagram ($|U|/t,n$) for the $s$-wave solution of 
the negative-$U$ Hubbard model in two dimensions ( with on-site attraction and 
nearest-neighbor hopping); 
(b) Phase diagram ($\epsilon_{o}$,$k_{F}^{2}/(2m)$) for a contact potential in 
two dimensions (as obtained from the available analytic solution).}
\end{figure}

It is interesting to compare the ``phase diagram'' of Fig.~3a for the 
two-dimensional negative-$U$ Hubbard model, with the ``phase diagram'' 
($\epsilon_{o}$,$k_{F}^{2}/(2m)$) (in units of $(m a_{o}^{2})^{-1}$, where 
$a_{o}$ is again an arbitrary unit of length) for the contact potential in the 
two-dimensional continuum case, for which the analytic solution is also 
available.\cite{Randeria-90,MPS}
This ``phase diagram'' is shown in Fig.~3b, where now the boundary curves with  
$k_{F} \xi_{pair} = (2 \pi,1/ \pi)$ correspond to straight lines.
[We have verified numerically that the solutions to the coupled equations
(\ref{Delta}) and (\ref{mu}) for the negative-$U$ Hubbard model and for the contact
potential in two dimensions coincide within a few percent when $n \lapprox 0.1$,
even for values of $|U|/t$ of the order of some units.]
Note that the \emph{absence of a threshold\/} for the occurrence of a bound state
in two dimensions makes the density-induced BCS-BE crossover possible even for a 
short-range potential, in contrast to the behavior for a three-dimensional 
contact potential obtained previously (cf. Fig.~1).

A comment on the nature of the bosonic limit for large values of $|U|/t$ when 
approaching half filling in Fig.~3a is in order at this point.
For the continuum case, the limit of a ``dilute'' gas of composite bosons can be
reached for any particle density, insofar as the range of the residual bosonic
interaction (or, equivalently, the size of the composite bosons) vanishes for large
values of the fermionic interaction strength (barring the instability problem
mentioned at the end of Section II).
In the lattice case, instead, the lattice spacing provides an additional length
scale in the problem, which makes it possible to depart from the ``dilute'' gas
limit irrespective of the size of the composite bosons.
It is, in fact, the ``overlap'' of the centers of mass of the composite bosons,
which is forced by increasing the density on the lattice, to make the usual ``dilute''
gas condition $k_{F} a << 1$ (as defined for the continuum model) no longer 
representing a ``dilute'' gas situation in the lattice case.
For such a high-density gas of composite bosons, therefore, the underlying fermionic
degrees of freedom are expected to become significant again.\cite{NSR}

To make these arguments more quantitative, let us consider the following boson-like
operator:

\begin{equation}
b_{o}^{\dagger} \, = \, \sum_{{\mathbf k}} \, g({\mathbf k}) \, 
c^{\dagger}_{{\mathbf k} \uparrow} \, c^{\dagger}_{-{\mathbf k} \downarrow} \,
                                                              \label{bosonic-b}
\end{equation}

\noindent
where $c^{\dagger}_{{\mathbf k} \sigma}$ creates a fermion with wave vector
${\mathbf k}$ and spin $\sigma$, $g({\mathbf k})$ represents the pair wave function,
and the sum over ${\mathbf k}$ is limited to the Brillouin zone in the lattice case.
The ensuing commutator

\begin{equation}
[b_{o},b_{o}^{\dagger}] \, = \, \sum_{{\mathbf k}} \, |g({\mathbf k})|^{2} \,
\left( 1 \, - \, n_{{\mathbf k} \uparrow} \, - \, n_{-{\mathbf k} \downarrow} \right)
                                                          \label{commutator}
\end{equation}

\noindent
with $n_{{\mathbf k} \sigma} = c^{\dagger}_{{\mathbf k} \sigma} c_{{\mathbf k} \sigma}$,
can be regarded a $c$-number provided $\langle n_{{\mathbf k} \sigma} \rangle << 1$
over the relevant set of states.\cite{KK}
The normalization condition

\begin{equation}
\sum_{{\mathbf k},\sigma} \, \langle n_{{\mathbf k} \sigma} \rangle \, = \, N
                                                                        \label{N}
\end{equation} 

\noindent
($N$ being the total number of fermions), implies that 
$\langle n_{{\mathbf k} \sigma} \rangle \ge n/2$ over a \emph{finite\/} region of 
${\mathbf k}$ space, whenever the range of the sum over ${\mathbf k}$ remains limited 
(as for the lattice case).
This restriction, in turn, implies that the commutator (\ref{commutator}) cannot be
considered as a $c$-number, as soon as $n$ is an appreciable fraction of the
unity.
Note that the size of the composite bosons does not enter the above argument. 

In summary, we have argued on quite general grounds that, to reach a satisfactory 
bosonic limit in the lattice case, the condition $k_{F} \xi_{pair} < 1/ \pi$ valid
in the continuum case has to be supplemented by the condition $n <<1$.

\subsection{$d$-wave solution}

When $V<0$ and $U>0$ in Eq.~(\ref{V-Hubbard}), a $d$-wave solution of the type
$\Delta({\mathbf k}) = \Delta_{1} (\cos k_{x} - \cos k_{y})$ can be considered.
In this case, we take either a nearest-neighbor dispersion relation
$\xi({\mathbf k}) = - 2 t (\cos k_{x} + \cos k_{y}) + n(U+4V)/2 - \mu$
as before ($t-V$ model), or a second- and third-neighbor dispersion relation
$\xi({\mathbf k}) = 4 t' \cos k_{x} \cos k_{y}
+ 2 t'' (\cos 2k_{x} + \cos 2k_{y}) + n(U+4V)/2 - \mu$ with $t' >0$ and $t'' >0$
($t'-t''-V$ model).
In both cases we have included the Hartree shift.\cite{footnote-3}
The latter dispersion relation favors the formation of bound pairs with $d$-wave 
symmetry at low density,\cite{Dagotto} the critical value $V_{c}$ (for a 
bound state to appear in the two-body problem) vanishing when $t''<0.5 t'$.
For the former dispersion relation limited to nearest-neighbor sites, on the other
hand, a finite value of $V_{c}$ occurs also for the $d$-wave solution at zero
density.\cite{Micnas}

As already mentioned, the two different single-particle dispersion relations that
we have adopted are meant to mimic the low-energy electronic band structure for
the cuprates in different ranges of the doping level $\delta$.
Specifically, the ($t',t''$)-dispersion captures the small-arc features of the
Fermi surface detected in underdoped cuprates,\cite{Ino,Marshall} for which we
can interpret $n \sim \delta \lapprox 0.15$.
The $t$-dispersion, on the other hand, reproduces the main features of the large 
Fermi surface and its doping dependence for nearly optimally doped 
cuprates,\cite{Saini} provided one interpretes $n \sim 1 - \delta$ with 
$0.15 \lapprox \delta \lapprox 0.30$.

The  ``phase diagrams'' for the $t-V$ and $t'-t''-V$ models are shown, respectively,
in Figs.~4a and 4b, where the boundaries between different regions have been 
identified as in Fig.~3a.\cite{footnote-5}
Note in Fig.~4a the occurrence of a finite critical value $V_{c}$ ($V_{c}/4t = 1.83$),
past which the BE region develops for $n=0$ ($V_{c}$ vanishes instead in Fig.~4b).
We also mention that the BCS-like region, corresponding to small values of $|V|$ in 
Figs.~4a and 4b, consistently supports nonvanishing values of the gap $\Delta_{1}$, 
contrary to a recent statement for the $t-V$ model.\cite{Hertog}
Note finally that the density-induced BCS-BE crossover is allowed for the $t'-t''-V$
model but not for the $t-V$ model.
This finding is, in turn, consistent with our previous assertions that the $t'-t''-V$ 
model might be relevant to the underdoped range of the cuprate superconductors and 
that the BCS-BE crossover scenario might apply to that range.

A comment on the \emph{additional\/} region at the upper-right corner of the
``phase diagrams'' in Figs.~4a and 4b, which is delimited by a long-dashed line and 
has been identified as ``correlated bosons'' (CB), is in order.

\begin{figure}
\narrowtext
\epsfxsize=8truecm
\epsfbox{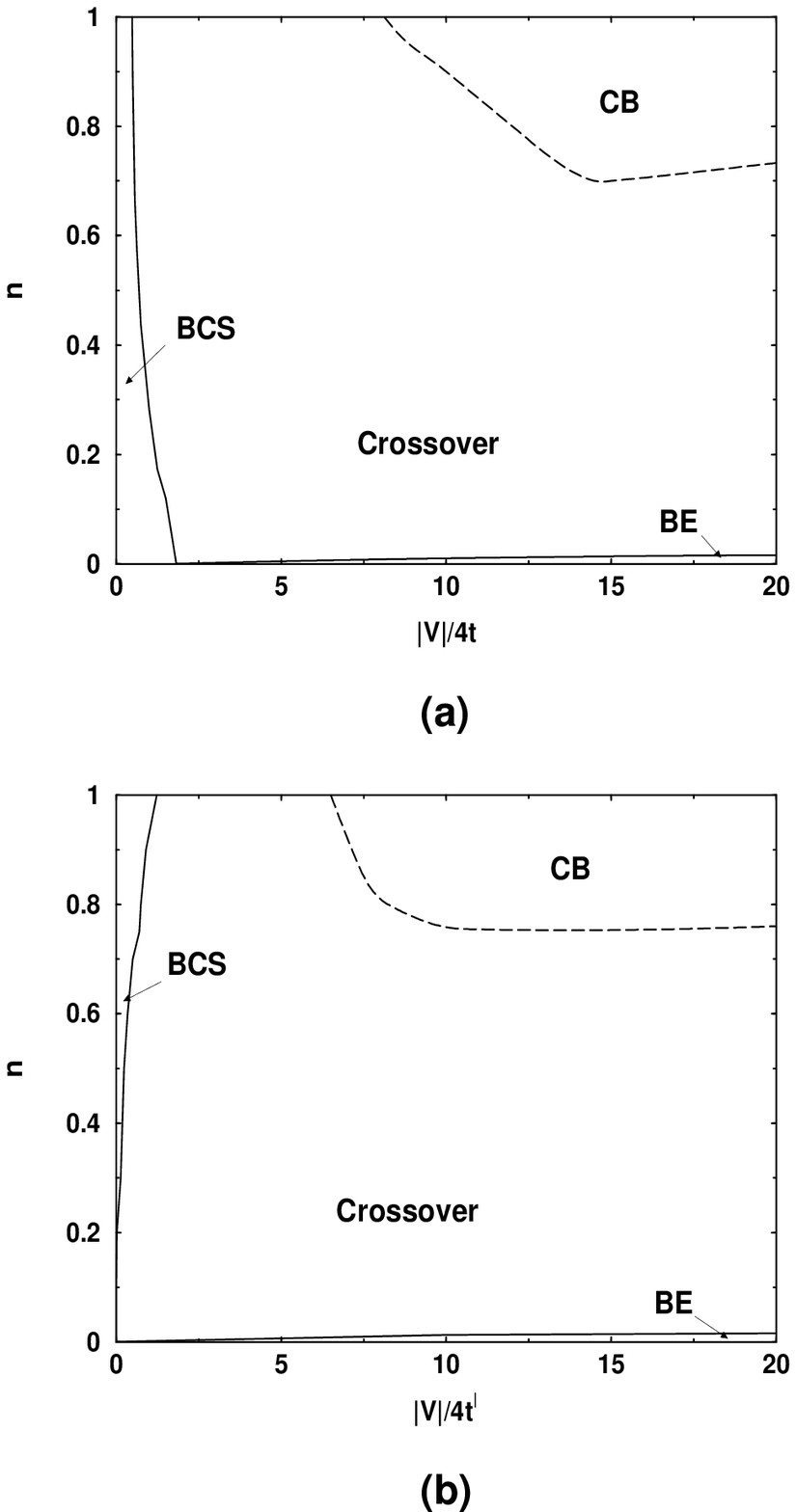}
\vspace{0.4truecm}
\caption{Phase diagram ($|V|/4t,n$) for the $d$-wave solution of the 
extended Hubbard model with attraction between nearest-neighbor sites in two 
dimensions, considering (a) nearest-neighbor or (b) second- and third-neighbor 
hopping.}
\end{figure}

This region is associated with a peculiar behavior of $\xi_{pair}$ for the $d$-wave
solution as a function of $|V|$ when $n$ approaches half filling, in the sense that 
$\xi_{pair}$ does not show a monotonic decrease for increasing $|V|$ and converges 
asymptotically (when $|V| \rightarrow \infty$ and $n<1$) to a finite value, which is 
larger than the lattice spacing.

This asymptotic value increases with $n$ and eventually diverges as $n \rightarrow 1$, 
making thus the product $k_{F} \xi_{pair}$ arbitrarily large.
[This peculiar behavior is absent for the $s$-wave solution discussed previously,
for which $\xi_{pair}$ is instead a monotonically decreasing function of $|V|$ for 
any given $n$. Consistently, the CB region is missing in Fig.~3a.]
On physical grounds, the divergence of $\xi_{pair}$ cannot be attributed either to 
the system converging to a BCS-like regime with large overlapping Cooper pairs or 
to the size of the composite bosons becoming infinitely large.
Rather, the divergence of $\xi_{pair}$ is due to the establishing of 
\emph{quasi-long-range-order\/} correlations among the composite bosons, which reside
individually on nearest-neighbor sites.
Under these circumstances, $\xi_{pair}$ weights preponderantly the correlation between 
opposite-spin fermions belonging to \emph{different\/} composite bosons, rather than 
the usual intra-boson correlation.
Accordingly, for the $d$-wave solution when $n \rightarrow 1$, $\xi_{pair}$ no longer 
represents the radius of the composite bosons in the strong-coupling limit.

To make the above argument more quantitative, we recall the original definition of
$\xi_{pair}$ in terms of the two-particle correlation function:\cite{PS-94,PS-96}

\begin{equation}
\xi_{pair}^{2} \, = \, \frac{\int \! d {\mathbf r} \,\,
                  g_{\uparrow,\downarrow}({\mathbf r}) \, {\mathbf r}^{2}}
                            {\int \! d {\mathbf r} \,\, 
                  g_{\uparrow,\downarrow}({\mathbf r})} \,\,  ,         \label{xi-pair}
\end{equation}

\noindent
where

\begin{eqnarray}
g_{\uparrow,\downarrow}({\mathbf r}) \, &=&\, \frac{1}{n^{2}}\, \langle
\psi^{\dagger}_{\uparrow}({\mathbf r}) \psi^{\dagger}_{\downarrow}({\mathbf 0})
\psi_{\downarrow}({\mathbf 0}) \psi_{\uparrow}({\mathbf r}) \rangle \, - \, 
\frac{1}{4} \,\nonumber\\
&=&\,\frac{1}{n^{2}}\, \left| \langle \Phi 
| \psi^{\dagger}_{\uparrow}({\mathbf r}) \psi^{\dagger}_{\downarrow}({\mathbf 0}) |
\Phi \rangle \right|^{2}                             \label{pair-correlation-function}
\end{eqnarray}

\noindent
is the pair-correlation function for opposite spin fermions (the last expression
holding specifically for the BCS ground state $|\Phi\rangle$).
From this definition, it appears evident that, in principle, 
$g_{\uparrow,\downarrow}({\mathbf r})$ does not distinguish between opposite-spin
fermions belonging to the \emph{same\/} pair or to \emph{different\/} pairs.
In practice, in the strong-coupling limit $g_{\uparrow,\downarrow}({\mathbf r})$
represents (the square of) the pair wave function whenever no correlation is 
established among the composite bosons; in this case, $\xi_{pair}$ tends to the
bound-state radius, as one verifies for the $s$-wave solution.
In the case that a definite correlation is established among the composite bosons,
on the other hand, $g_{\uparrow,\downarrow}({\mathbf r})$ spreads over a large (and
even infinite) number of lattice sites, and $\xi_{pair}$ increases (and eventually
diverges) accordingly. 
In this case, $g_{\uparrow,\downarrow}({\mathbf r})$ embodies the correlation among
different composite bosons and is totally unrelated to the pair wave function.

The occurrence of this novel feature for the $d$-wave solution is evidenced in 
Fig.~5a, where the amplitude

\begin{equation}
\phi({\mathbf R}_{n}) \, = \, \frac{1}{{\mathcal N}} \, \sum_{{\mathbf k}}^{BZ} \, 
\exp\{-i {\mathbf k} \cdot {\mathbf R}_{n}\} \, \frac{\Delta({\mathbf k})}{2 E({\mathbf k})}
                                                                \label{amplitude} 
\end{equation}

\noindent
is reported for $|V|/t>>1$ and $n=1$ over a grid of lattice sites ${\mathbf R}_{n}$ 
(${\mathcal N}$ being the total number of sites); in this way, 
$|\phi({\mathbf R}_{n})|^{2}$ represents the lattice version of 
$g_{\uparrow,\downarrow}({\mathbf r})$ given by Eq.~(\ref{pair-correlation-function}).
We see from this figure that the amplitude (\ref{amplitude}) develops clear structures
near the lattice diagonals, thus establishing definite correlations among 
opposite-spin fermions belonging to different composite bosons.\cite{footnote-p-h}

\begin{figure}
\narrowtext
\epsfxsize=8truecm
\epsfbox{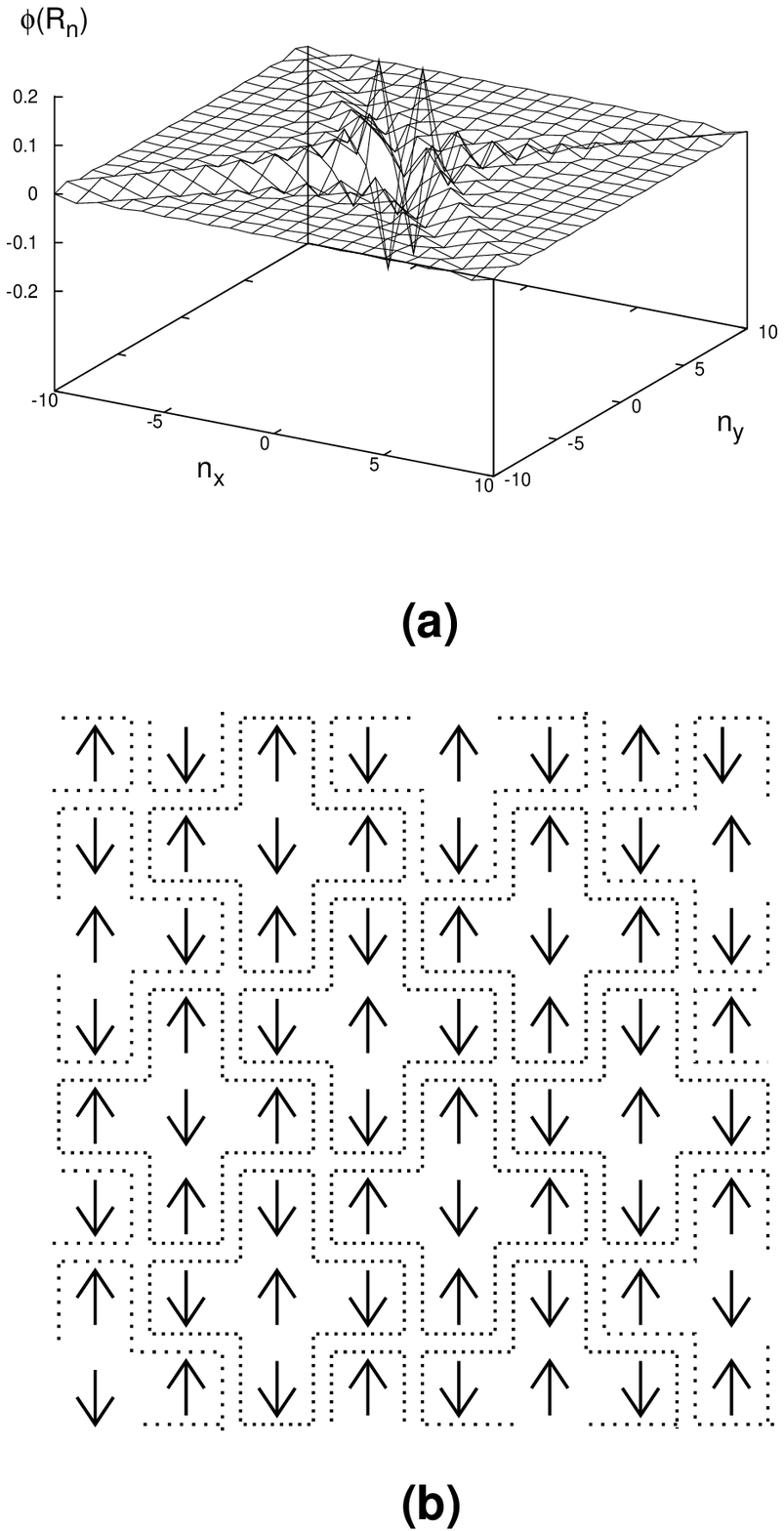}
\vspace{.4truecm}
\caption{(a) $d$-wave amplitude $\phi({\mathbf R}_{n})$ over the sites
${\mathbf R}_{n}$ of a two-dimensional square lattice, obtained for the $t-V$ model
when $|V|/t>>1$ and $n=1$; (b) Pictorial representation of the ordering of the
composite bosons with $d$-wave symmetry on the two-dimensional square lattice
when $|V|/t>>1$ and $n=1$.}
\end{figure}

This behavior can also be checked analytically, making use of the fact that at
half filling $\mu$ equals the Hartree shift $(U + 4V)/2$ and thus the single-particle 
dispersion $\xi({\mathbf k})$ is negligible in comparison to $|\Delta({\mathbf k})|$ 
when $|V|/t>>1$ (barring values of ${\mathbf k}$ for which $\Delta({\mathbf k})$ 
vanishes by symmetry).
In this case

\begin{equation}
\frac{\Delta({\mathbf k})}{2 E({\mathbf k})} \, = \, 
\frac{\Delta({\mathbf k})}{2 |\Delta({\mathbf k})|} \, = \, \frac{1}{2} \,
\mathrm{sgn} \left( \cos k_{x} - \cos k_{y} \right)                 \label{sign}
\end{equation}

\noindent
in Eq.~(\ref{amplitude}) alternates sign according to the $d$-wave symmetry, and 
its lattice Fourier transform

\begin{equation}
\phi({\mathbf R}_{n}) \, = \, \frac{1}{\pi^{2}} \, \frac{\left[1-(-1)^{n_{x}+n_{y}}\right]}
                       {n_{x}^{2}-n_{y}^{2}}                      \label{analytic}
\end{equation}

\noindent
(with ${\mathbf R}_{n}=(n_{x},n_{y})$ in units of the lattice spacing) decays as a
power law for increasing distance when $n_{x}+n_{y}$ is an odd integer, while it 
vanishes identically when $n_{x}+n_{y}$ is an even integer.
[In contrast, for the $s$-wave solution $\Delta({\mathbf k})/E({\mathbf k})$
tends to a constant value when $|V|/t>>1$ and $n=1$, and the corresponding Fourier 
transform is a Kronecker delta $\delta_{{\mathbf R}_{n},{\mathbf 0}}$.]
The ensuing picture bears strong resemblance with an antiferromagnetic ordering on
a square lattice, with opposite spins alternating over the two interpenetrating
sublattices in which the square lattice can be partitioned.
The ordering of the composite bosons (in the strong-coupling limit when approaching
half filling), as envisaged from the above considerations, is shown schematically in 
Fig.~5b.
Note that the \emph{quasi-long-range order\/} associated with the algebraic decay in
Eq.~(\ref{analytic}) corresponds to a divergent $\xi_{pair}$, even though it
does not represent a true long-range antiferromagnetic order of the spins.
Physically, such a strong correlation among fermions with opposite spins stems from
the original fermionic attraction between opposite-spin fermions residing on
nearest-neighbor sites, which corresponds to the form (\ref{V-Hubbard}) of the
potential.
It is for these reasons that we have identified the region in the upper-right corner 
of Figs.~4a and 4b (delimited by a long-dashed line) as ``correlated bosons'' 
rather than BCS-like, even though $k_{F} \xi_{pair}$ becomes definitely larger than 
$2 \pi$ in this region.\cite{footnote-6}

Another relevant difference of the $d$- from the $s$-wave solution, 
is the fact that in the BE region of Figs.~4a and 4b ($-2\mu$) does \emph{not\/} reduce 
to the bound-state energy $\epsilon_{o}$ of the associated two-body problem, unless
$n=0$ strictly.
This additional feature of the $d$-wave solution can also be associated with the
augmented correlation among the composite bosons with increasing $n$, as discussed
previously.
Such a correlation makes, in fact, the energy required to extract two fermions from
the system different from the energy required to break up a \emph{single\/} composite
boson in isolation, owing to the additional correlation energy among the composite 
bosons.

The relevance of the correlation energy among the composite bosons is also suggested
by the behavior of the ``bosonic'' chemical potential $\mu_{B} = 2\mu + \epsilon_{o}$ 
in the BE region (delimited by $n \le 0.016$, irrespective of the single-particle
dispersion). 
In this low-density region one finds $\mu_{B}=(U+2V)n$.
When $0<U<2|V|$, $\mu_{B}$ is negative, corresponding to an effective average
\emph{attraction\/} between the composite bosons.
The compressibility of the system is also negative at this order, thus indicating
a tendency toward phase separation.\cite{Micnas}
When $U>2|V|$, on the other hand, $\mu_{B}$ and the compressibility are both
positive; in this case, the effective bosonic average interaction would be 
\emph{repulsive\/}, with an increasing repulsion between the composite bosons
as $|V|$ (and $U$) increases.
This situation has to be contrasted with the negative-$U$ Hubbard model, for 
which the composite bosons become asymptotically free as 
$|U| \rightarrow \infty$.\cite{footnote-3}

\section{Concluding Remarks}

In this paper we have examined how the particle density influences 
the BCS-BE crossover, by analyzing several types of fermionic interaction potentials
both in three dimensions (continuum case) and in two dimensions (continuum and lattice 
cases).
We have reached the conclusion that the finite range of the potential as well as the 
absence of a threshold (for a bound state to occur in the associated two-body problem) 
favor quite generally the density-induced crossover.

In particular, for the continuum case we have verified that in three dimensions 
(where a finite threshold exists in the two-body problem) it is the finite range of 
the potential to make the density-induced crossover possible, while in two dimensions 
the absence of a threshold suffices to the purpose even when a zero-range potential 
is considered.
In addition, for the continuum case we have argued that the density-induced crossover
(for a finite-range potential) occurs at arbitrary values of the interaction strength,
owing to the fact that, in this case, the particle density can be augmented without 
bound.

For the lattice case, we have considered two dimensions only and investigated,
alternatively, the absence or presence of a threshold, depending on the symmetry
of the gap parameter and on the shape of the single-particle dispersion relation.
We have found that the absence of a threshold definitely favors the density-induced
BCS-BE crossover, and that spreading the range of the potential on the lattice makes
the density-induced BCS-BE crossover to occur over a wider density range.
In the presence of a threshold (at least for the specific potential extending over 
nearest-neighbor sites that we have considered), on the other hand, the density-induced
BCS-BE crossover turned out not to be possible, owing essentially to the absence of
the reentrant shape of the curve delimiting the BCS-like region.
This situation contrasts our finding for the continuum case in the presence
of a threshold.

In this respect, an important difference between the continuum and lattice cases
appears to be the occurrence of an intrinsic \emph{upper value for the density \/} in 
the lattice case, at least when one considers a simple band only.
In this case, the Fermi wave vector $k_{F}$ cannot exceed an upper bound of the
order $\pi$ (in units of the inverse of the lattice constant).
For the nearest-neighbor interaction which we have considered in the lattice case,
the characteristic wave vector $k_{o}$, too, can be taken of the order $\pi$ and
the condition $k_{F} >> k_{o}$ (which in the continuum case was associated with the
reentrant shape) cannot be satisfied.
Consequently, no reentrant shape of the curve delimiting the BCS-like region is
expected for the lattice case.
The occurrence of the reentrant shape, however, may not \emph{a priori\/} be excluded 
when considering an interaction potential extending over distant neighbors in the 
lattice, thus decreasing $k_{o}$ accordingly.
In any case, the occurrence of an intrinsic upper bound on the density (and thus on 
$k_{F}$) will make the BCS-like region to disappear for large enough interaction 
strength, thus preventing the density-induced BCS-BE crossover.

We have also found that the presence of an upper bound on the density in the lattice
case and the finite size of the composite bosons for the $d$-wave solution induce
definite correlations among the composite bosons, giving rise to a magnetic correlated
state near half filling.
The simultaneous presence of superconducting off-diagonal long-range order
\emph{and\/} of antiferromagnetic correlations without long-range order, which we
have found for the $d$-wave solution, represents \emph{per se\/} an appealing
result, in the light of the current phenomenology of the cuprate superconductors.
The interesting physics that has emerged for the $d$-wave solution in the lattice
case may then provide a general framework for future investigations on the BCS-BE
crossover, for instance, considering finite-temperature effects and going beyond
the mean-field approximation.
In this respect, facing a proper treatment of the residual boson-boson interaction
in the BE region seems to be unavoidable, not only in the three-dimensional continuum 
case (as shown already in Ref.~\cite{Pi-S-98}), but especially in the two-dimensional
lattice case for the $d$-wave solution, where the boson-boson interaction gets 
considerably enhanced.

Finally, we point out that the need of a reduced spatial dimensionality and of a 
finite range of the potential, which we have found for the density-induced
BCS-BE crossover to occur, matches the generic features observed for the cuprate
superconductors and strengthen accordingly the BCS-BE crossover as a possible 
scenario for describing the evolution from overdoped to underdoped cuprates.

\section*{Acknowledgments}

We are indebted to C. Castellani for discussions.
One of us (P.P.) gratefully acknowledges receipt of a postdoctoral research 
fellowship from the Italian INFM under contract PRA-HTCS/96-98.


\end{multicols}


\begin{thebibliography}{99}

\bibitem{NSR} P. Nozi\`{e}res and S. Schmitt-Rink, J. Low. Temp. Phys. {\bf 59}, 
195 (1985). 

\bibitem{Randeria-93} C.A.R. S\'a de Melo, M. Randeria, and J.R. Engelbrecht,
Phys. Rev. Lett. {\bf 71}, 3202 (1993).

\bibitem{Haussmann} R. Haussmann, Z. Phys. B{\bf 91}, 291 (1993).

\bibitem{PS-96} F. Pistolesi and G.C. Strinati, Phys. Rev. B{\bf 53}, 15168 (1996).

\bibitem{Zwerger} S. Stintzing and W. Zwerger, Phys. Rev. B{\bf 56}, 9004 (1997).

\bibitem{Levin} B. Jank\'o, J. Maly, and K. Levin, Phys. Rev. B{\bf 56}, R11407
(1997).

\bibitem{Uemura} Y.J. Uemura \emph{et al.\/}, Phys. Rev. Lett. {\bf 66}, 2665 (1991);
Nature {\bf 352}, 605 (1991).

\bibitem{footnote-1} It is, in principle, possible for the attractive potential to
vary with carrier density. 
[This effect can be found, e.g., near a Quantum Critical Point driven by the carrier
density itself (C. Castellani, C. Di Castro, and M. Grilli, Z. Phys. B{\bf 103},
137 (1997)).]
Knowledge, however, of the form of the interparticle potential and of its dependence 
on carrier density rests on an explicit microscopic theory, which is beyond the scope 
of the present crossover approach. By this approach, in fact, the interparticle 
potential is parametrized in a rather simple form, independent from the particle 
density. 

\bibitem{Micnas} R. Micnas, J. Ranninger, and S. Robaszkiewicz, Rev. Mod. Phys. 
{\bf 62}, 113 (1990).

\bibitem{Randeria-98} J.R. Engelbrecht, A. Nazarenko, M. Randeria, and E. Dagotto,
Phys. Rev. B{\bf 57}, 13406 (1998). 

\bibitem{MPS} M. Marini, F. Pistolesi, and G.C. Strinati, Eur. Phys. J. B{\bf 1}, 
151 (1998).

\bibitem{PS-94} F. Pistolesi and G.C. Strinati, Phys. Rev. B{\bf 49}, 6356 (1994).

\bibitem{Perali} G. Varelogiannis, A. Perali, E. Cappelluti, and L. Pietronero,
Phys. Rev. B{\bf 54}, R6877 (1996).

\bibitem{footnote-2} Quite generally, it can be shown that the curves 
$k_{F} \xi_{pair} = \mathrm{constant}$: (i) fill up the plane 
($|V|/V_{c}$,$k_{F}/k_{o}$); (ii) never intersect each other; and (iii) all depart
from the point ($|V|/V_{c}=1$,$k_{F}=0$).

\bibitem{Pi-S-98} P. Pieri and G.C. Strinati, cond-mat/9811166.

\bibitem{Ropke} G. R\"{o}pke \emph{et al.\/}, Phys. Rev. Lett. {\bf 80}, 3177 (1998).

\bibitem{Ino} A. Ino \emph{et al.\/}, cond-mat/9902048.

\bibitem{Saini} N.L. Saini \emph{et al.\/}, Phys. Rev. Lett. {\bf 79}, 3467 (1997).

\bibitem{Levin-98} Q. Chen, I. Kosztin, B. Jank\'{o}, and K. Levin, Phys. Rev. 
B{\bf 59}, 7083 (1999).

\bibitem{Hertog} B.C. den Hertog, cond-mat/9808051.

\bibitem{Dagotto} A. Nazarenko, A. Moreo, E. Dagotto, and J. Riera, Phys. Rev. 
B{\bf 54}, R768 (1996).

\bibitem{footnote-3} For the negative-$U$ Hubbard model, the Hartree
shift $nU/2$ makes the chemical potential to coincide with $-\epsilon_{o}/2$ in
the bosonic (strong-coupling) limit, \emph{irrespective\/} of the value of $n$
(in such a way that the bosonic chemical potential $\mu_{B} = 2\mu + \epsilon_{o}$
vanishes in the limit - cf. the result for $\mu$ given in Ref.~\cite{Micnas}).
[Recall instead that for the continuum case the inclusion of the Hartree shift is
not required.]
Accordingly, the physical picture which emerges in the strong-coupling limit would 
correspond to a system of non-interacting composite bosons (with vanishing size),
although near half filling this elementary picture is not completely satisfactory
owing to the overlapping argument discussed in the text.
In any event, we shall consider the negative-$U$ Hubbard model as the 
\emph{reference model\/} for the BCS-BE crossover in the lattice case, akin to the 
contact potential in the continuum case.
For the alternative model considered in the lattice case (with an attraction 
between nearest-neighbor sites), the same procedure of adding the Hartree shift 
will consistently be adopted.

\bibitem{Randeria-90} M. Randeria, J.-M. Duan, L.-Y. Shieh, Phys. Rev. B{\bf 41}, 
327 (1990).

\bibitem{KK} A similar commutator has been considered by L.V. Keldysh and A.N.
Kozlov [Sov. Phys. JETP {\bf 27}, 521 (1968)] in the context of the Bose 
condensation of excitons in semiconductors.

\bibitem{Marshall} D.S. Marshall \emph{et al.\/}, Phys. Rev. Lett. {\bf 76}, 4841
(1996).

\bibitem{footnote-5} For the $t'-t''-V$ model the density of states is no longer 
symmetric about half filling. Nevertheless, in Fig.~4b we have considered 
$0 \le n \le 1$ only, because the remaining values of $n$ are of no physical 
interest.

\bibitem{footnote-p-h} Although the $t'-t''-V$ model lacks the particle-hole
symmetry of the $t-V$ model, when $|V|$ is large enough the kinetic term of the
Hamiltonian becomes immaterial and the $t'-t''-V$ model behaves like the $t-V$
model. The results shown in Fig.~5a hold, therefore, for both models.

\bibitem{footnote-6} It can be verified that the quasi-long-range-order state with
antiferromagnetic correlations discussed in the text (which was obtained from the
BCS state in the limit $|V| \rightarrow \infty$ and $n \rightarrow 1$) has higher
energy than the N\'{e}el state (which is an eigenstate of the Hamiltonian in this
limit and exhibits true long-range order).
Let $H = - |V| \sum_{<i,j>} n_{i,\uparrow} n_{j,\downarrow} \,$, where $<i,j>$ runs
over all pairs of nearest-neighbor sites and the kinetic energy has been neglected 
in the limit.
It can then be verified that $\langle H \rangle$ is given by 
$- {\mathcal N} (1 + 16/\pi^{4}) |V|$ for the BCS state and by $- {\mathcal N} 2 |V|$
for the N\'{e}el state.
This difference can be ascribed to the fact that the BCS state orders the composite
bosons by condensing them in momentum space, while a more appropriate description
should correspond to the composite bosons being ordered in real space (as depicted
in Fig.~5b).
An improved wave function (over and above the BCS state) is thus required to describe 
the composite bosons in the limit $|V| \rightarrow \infty$ and $n \rightarrow 1$.
 
\end{thebibliography}
\end{document}